\documentclass{new_tlp}
\usepackage{mathptmx}
\usepackage{mathtools}
\newtheorem{definition}{Definition}

\usepackage{url}
\usepackage{listings}
\usepackage{amsfonts}

\usepackage{graphicx}
\usepackage{pgfplots}
\definecolor{success}{RGB}{70, 210, 50}
\definecolor{failed}{RGB}{220, 30, 30}
\usepackage{subcaption}
\usepackage{hhline}
\usepackage{xspace}
\usepackage{multirow}
\usepackage{soul}

\newcommand{\naf}{\mathop{not}}
\newcommand{\penalt}{\mathit{Penalty}}

\newcommand{\reviewfix}[1]{\textcolor{black}{#1}}
\newcommand\nCite[1]{\citeANP{#1} \shortcite{#1}}

\newtheorem{example}{Example}[section]
\newtheorem{theorem}{Theorem}[section]

  \title[Theory and Practice of Logic Programming]
        {Unit Testing in ASP Revisited: \\ Language and Test-Driven Development Environment}

  \author[G. Amendola, T. Berei, G. Mazzotta, F. Ricca]
         {GIOVANNI AMENDOLA$^1$ \and TOBIAS BEREI$^2$ \and GIUSEPPE MAZZOTTA$^1$ \and FRANCESCO RICCA$^1$\\
         $^1$University of Calabria, Rende, Italy\\
         \email{firstname.lastname@unical.it}\\
         $^2$University of Applied Sciences Upper Austria, Campus Hagenberg\\
         \email{tobias.berei@students.fh-hagenberg.at} \\ \email{bereitobias@hotmail.com}}

\jdate{March 2021}
\pubyear{2021}
\pagerange{\pageref{firstpage}--\pageref{lastpage}}
\doi{S1471068401001193}

\begin{document}

\label{firstpage}

\maketitle

  \begin{abstract}
Unit testing frameworks are nowadays considered a best practice, included in almost all modern software development processes, to achieve rapid development of \textit{correct} specifications. Knowledge representation and reasoning paradigms such as Answer Set  Programming (ASP), that have been used in industry-level applications,  are not an exception.
Indeed, the first unit testing specification language for ASP was proposed in 2011 as a feature of the ASPIDE development environment. 
Later, a more portable unit testing language was included in the LANA annotation language.
In this paper we revisit both languages and tools for unit testing in ASP.
We propose a new unit test specification language that allows one to inline tests within ASP programs, and 
we identify the computational complexity of the tasks associated with checking the various program-correctness assertions. 
Test-case specifications are transparent to the traditional evaluation, but can be interpreted by a specific testing tool.
Thus, we present a novel environment supporting test driven development of ASP programs.

  \end{abstract}



\section{Introduction}
Answer Set Programming (ASP)~\cite{DBLP:journals/cacm/BrewkaET11} is a well-known \reviewfix{logic-based} formalism developed in the area of knowledge representation and reasoning.
ASP combines a purely declarative language based on the stable models semantics~\cite{DBLP:journals/ngc/GelfondL91} with efficient implementations~\cite{DBLP:journals/aim/LierlerMR16}.
ASP is known to be suited for rapid prototyping of complex reasoning tasks and has been effectively used to solve a number of both academic and real-world applications of AI~\cite{DBLP:journals/aim/ErdemGL16}.
ASP allows encoding complex computational problems often in an easier and more compact way than mainstream (imperative) programming languages.
For instance, the classical NP-complete problem of 3-Colorability is encoded in ASP using only two rules.
Nonetheless, it is also easy to write incorrect ASP programs, which seem correct (often due to a misleading interpretation of rules based on intuition) but do not work as expected. 

To speed up the development of correct and robust programs, many modern software development processes support  some \textit{Test-Driven Development} (TDD)  best practices \cite{DBLP:conf/xpu/FraserBCMNP03} such as \textit{unit testing}~\cite{Beck2002}.
In TDD the following sequence of actions is repeated while developing a new program~\cite{Beck2002}:
(i) add a test that defines a function or improvements of a function, which should be very succinct, so that the developer focuses on the requirements before writing the code; (ii) run all tests and see if the new test fails.
This confirms that the newly introduced tests indeed bring an improvement in detecting errors;
(iii) write the code (maybe not perfect); (iv) run all tests, to become confident that the new code meets the test requirements, and does not introduce bugs; (v) refactor code to improve it while accommodating the new features. 

Tests drive the development because the program is considered improved only if it passes new tests, while the repeated execution of tests allows to find problems early in the development cycle and to isolate the incorrect behavior more easily. 
This intuition has been subject to further empirical studies which proved that programmers who wrote more tests tended to be more productive~\cite{DBLP:journals/tse/ErdogmusMT05} and evidenced the superiority of the TDD practice over the traditional test-last approach or testing for correctness approach~\cite{DBLP:books/sp/Madeyski10}.
Two important best practices of TDD are the following: (i) concentrate on testing possibly small modules/functions/blocks of code; and (ii) adopt frameworks and tools to make the process of testing the program automatically. 
In the resulting software testing method, complex programs are split into \textit{units}, which are tested in isolation providing (usually) small inputs and checking whether the expected outputs are computed. 
Tests for program units (i.e., \textit{unit tests}) are directly derived from software requirements and often created before (or while) implementing the corresponding feature. 
The software is improved to pass the new tests, only. 

TDD development practices are nowadays a standard technique used by expert programmers and development teams all over the world, no matter the programming language used.
TDD development practices have been already applied also to many AI formalisms, as it is witnessed by the proposals in the literature, such as testing for Description Logics~\cite{DBLP:conf/ontobras/BezerraF17}, and Constraint Programming~\cite{DBLP:conf/cp/LazaarGL10}. 
In particular, TDD development practices have been applied  to ASP program development.
Indeed the first unit testing language for ASP has been introduced in 2011~\cite{DBLP:conf/inap/FebbraroLRR11} and was implemented in ASPIDE~\cite{DBLP:conf/lpnmr/FebbraroRR11}. 
Nonetheless, this solution presents some limitations from the perspectives of both the language for specifying unit tests and its implementation. 
Concerning the language, one drawback is the need for specifying unit tests in separate files concerning the program to test, which often is not very comfortable given that ASP programs might contain (comparatively) few lines of code.
Having the possibility to write tests \textit{together} with the source code, without interfering with the actual evaluation of the program itself, would provide clear advantages for the developer.
%
Subsequently~\cite{DBLP:journals/tplp/VosKOPT12}, a more versatile approach of specifying unit tests together with ASP programs was included in the LANA annotation language. 
However, the prototypical implementation of LANA, called ASPUnit, has never been included in a graphical development environment for ASP. A testing framework integrated into a development environment is comfortable for the developer, who can control the results in the same graphical environment she is using.


In this paper, we propose a new unit testing language for ASP, which comprises the key features of both the above-mentioned proposals, and we identify the computational complexity of the tasks associated with checking the various program-correctness conditions, which can be specified in unit tests. To the best of our knowledge, this is the first attempt to identify the complexity for all the tasks connected with unit testing in ASP.
The new language is annotation-based as LANA, so to allow the development of test cases inline with ASP code, and keep the assertion-based style for expressing test case conditions from the language of ASPIDE. 
Note that, being capable of defining and running tests together with the program has several clear advantages, such as promoting and exploiting programmer familiarity with the language (there is no need to know both ASP and another language such as python) and enables programmers to define declarative test case specifications.
The proposed annotation style is more similar to the JUnit framework for Java (from which both ASPIDE and LANA proposals were inspired), and should look more familiar to developers that are accustomed to XUnit style languages~\cite{Beck2002}.
Moreover, we present a novel web-based development  environment for ASP supporting test-driven development of ASP programs that features an integrated implementation of unit testing.

We also observe that there is one additional advantage in equipping a programming language with unit tests: they can be used to check and certify the behaviors of programs with respect to \textit{any} requirement of the application~\cite{DBLP:conf/xpu/FraserBCMNP03}.
For example, ethical or trustful behavior can be seen as a software requirement~\cite{DBLP:books/lib/Sommerville07}. 
Unit testing languages can be used to devise tests that check and certify that the behavior of the ASP program respects also application-specific ethical requirements.
Programs that are provided without tests of such type might result in (unwanted) violation of ethical requirements (besides potential bugs). 
In a sense, our contribution can lead to the development of ASP-based AI applications that can be certified with respect also to ethical requirements and other specific expected behaviors by reporting the outcome of executing properly-devised test suites.

\paragraph{Conference paper.} This work is an extended and revised version of the paper ``Testing in ASP: Revisited language and programming environment'' accepted for publication in the 17th edition of the European Conference on Logics in Artificial Intelligence~\cite{DBLP:conf/jelia/AmendolaBR21}.
Preliminary results were illustrated during the 35th International Conference on Logic Programming ICLP 2019~\cite{DBLP:journals/corr/abs-1909-07646}.
The current version of the contribution extends the conference paper by:
\begin{enumerate}
\item Providing the complexity analysis of the main unit testing tasks.
\item Extending the description of the implementation.
\item Presenting the results of an experimental analysis that validates the applicability and efficiency of the proposed approach.
\end{enumerate}

The investigation of computational complexity makes clear that testing an ASP program can be computationally very expensive (up to $\Pi_2^P$-c, cfr. Table~\ref{tab:Complexity}). This result not only fills a gap in the understanding of unit testing in ASP, but also enabled the development of a (more) efficient implementation, as evidenced empirically by the experimental analysis presented in this paper. Moreover, it serves as a valuable reminder to programmers that they should rely on the small scope hypothesis~\cite{DBLP:conf/kr/OetschPPST12} when crafting effective test cases. In other words, it is crucial to design small tests that cover the defects, while avoiding the use of large instances that may become impractical due to their high complexity. This understanding aids in promoting efficient and effective testing practices in ASP development.

\paragraph{Paper structure.} This paper is organized as follows. Section~\ref{sec:preliminaries} provides a preliminary introduction to ASP and introduces some basic notation; Section~\ref{sec:unit} introduces the new unit testing language by resorting to simple examples; Section~\ref{sec:complexity} reports a study of the computational complexity of the main computational tasks related to unit testing; Section~\ref{sec:env} describes the implementation of our programming environment; \reviewfix{Section~\ref{sec:experiments} reports an experimental evaluation aimed at demonstrate the effectiveness, but also the efficiency, of the proposed system in detecting bugs in common ASP problems.}
Section~\ref{sec:related} discusses related work; Section~\ref{sec:conclusion} reports our conclusive statements.

\section{Preliminaries on Answer Set Programming}\label{sec:preliminaries}

Let $\mathcal{P}$ be a set of predicates, $C$ of constants, and $V$ of variables. A \textit{term} is a constant or a variable. 
An atom $a$ of arity $k$ is of the form $p(t_{1},..., t_{k})$, where $p\in\mathcal{P}$ and $t_{1},..., t_{n}$ are terms. 
A \textit{disjunctive rule} $r$ is of the form
\begin{equation}
 a_{1}\vee\ldots\vee a_{l} \leftarrow b_{1},\ldots,b_{m}, \ not \
 c_{1},\ldots,\ not \ c_{n},
\label{eq:rule}
\end{equation}
\noindent where all $a_i$, $b_j$, and $c_k$ are atoms; $l,m,n\geq 0$ and $l+m+n >0$; $not$ represents
\textit{negation-as-failure}. 
The set $H(r)=\lbrace
a_{1},...,a_{l} \rbrace$ is the \textit{head} of $r$; $B^{+}(r)=\lbrace 
b_{1},...,b_{m} \rbrace$ and $B^{-}(r)=\lbrace
c_{1},\ldots,c_{n} \rbrace$ are
the \textit{positive body} and the \textit{negative body} of $r$,
respectively; and $B(r)=B^{+}(r)\cup B^{-}(r)$ is the \textit{body} of $r$.
A rule $r$ is \textit{safe} if each of its variables occurs in some positive 
body atom. 
A rule $r$ is a \textit{fact}, if $B(r)=\emptyset$ (we then omit
$\leftarrow$ from the notation);
a \textit{constraint} if $H(r)=\emptyset$;
\textit{normal} if $| H(r)| \leq 1$;
and \textit{positive} if $B^{-}(r)=\emptyset$.
A \textit{(disjunctive logic) program} $P$ is a
finite set of disjunctive rules. $P$ is called \textit{normal}
[resp.\ \textit{positive}] if each $r\in P$ is normal
[resp.\ positive]. 
Moreover, a program $P$ is \textit{head-cycle free}
if there is a level mapping  $\|.\|_h$ of $P$ such that for every rule $r$ of $P$: $(i)$ For any $l$ in $B^+(r)$, and for any $l'$ in $H(r)$,  $\| l \|_h \leq \| l' \|_h$; and
$(ii)$ For any pair $l, l'$ of atoms in $H(r)$,  $\| l \|_h\neq \| l' \|_h$.
In the paper, $At(r)$ denotes the set of all atoms in rule $r$, and $At(P)=\bigcup_{r\in P} At(r)$ is the set of all atoms in $P$.
We restrict attention to programs built on safe \reviewfix{rules only (to avoid well-known issues~\cite{DBLP:journals/tocl/LeonePFEGPS06}).}

The {\em Herbrand universe} of $P$, denoted by $U_{P}$, is the set of all 
constants appearing in $P$. If there are no constants in $P$, we
take $U_{P}=\{a\}$, where $a$ is an arbitrary constant.
The {\em Herbrand base} of $P$, denoted by $B_{P}$, is the set of all ground atoms that can be obtained
from the predicate symbols appearing in $P$ and the constants in $U_{P}$.
Given a rule $r$ of $P$, a {\em ground instance} of  $r$ is a rule obtained from $r$ by replacing every variable $X$ in $r$ by $\sigma (X)$, where $\sigma$  is a substitution mapping the variables occurring in $r$ to constants in $U_{P}$. 
The \textit{ground instantiation} of $P$, denoted by $ground(P)$, is the set of all the ground instances of the rules occurring in $P$. 

Any set $I\subseteq B_P$ is an \textit{interpretation}; 
\reviewfix{it \textit{satisfies} a rule $r \in ground(P)$ if $H_r \subseteq I$ (denoted as, $I \models H(r)$) or $B^+(r)\subseteq I$ and $B^-(r)\cap I = \emptyset$ (denoted as, $I \models B(r)$);
it is a \textit{model} of a program $P$ (denoted $I\models P$) if for each rule $r\in ground(P)$, $I \models r$.}
A model $M$ of $P$ is \textit{minimal} if no model $M'\subset M$
of $P$ exists.
We denote by $MM(P)$ the set of all minimal models of $P$. 
We write $P^I$ for the well-known 
\textit{Gelfond-Lifschitz reduct}~\cite{DBLP:journals/ngc/GelfondL91}
of $P$ w.r.t. $I$, that is, the set of rules $H(r) \leftarrow
B^{+}(r)$, obtained from rules $r\in ground(P)$ such that $B^-(r) \cap I= 
\emptyset$. We denote by $AS(P)$ the set of all {\em answer sets (or stable
models)} of $P$, that is, the set of all interpretations  $I$ such that
$I\in MM(P^I)$. 
We say that a program $P$ is \textit{coherent}, if $AS(P) \neq \emptyset$, otherwise, $P$ is \textit{incoherent}.


Finally, we recall a useful extension of the answer set semantics by the notion of {\em weak constraint}~\cite{DBLP:journals/tkde/BuccafurriLR00}.
A weak constraint $\omega$ is of the form:
\begin{equation}
	:\sim b_1,\ldots, \ b_m, \ \naf c_{1},\ldots, \ \naf c_n. \ [c@l],
\end{equation}
where $c$ and $l$ are nonnegative integers, representing a \textit{cost} and a \textit{level}, respectively.
Let $\Pi$ $=$ $P\cup W$ , where $P$ is a set of rules and $W$ is a set of weak constraints.
We call $M$ an answer set of $\Pi$ if it is an answer set of $P$.
We denote by $W(l)$ the set of all weak constraints at level $l$.
For every answer set
$M$ of $\Pi$ and any $l$, the \textit{penalty} of $M$ at level $l$, denoted by $\penalt_\Pi(M,l)$, is defined as
\reviewfix{$\sum_{\omega\in W(l), \ M\models B(\omega) } c$}. 
%
For any two answer sets $M$ and $M'$ of $\Pi$,
we say $M$ is \textit{dominated} by $M'$ if there is $l$ s.t.
$(i)$ $\penalt_\Pi(M',l) < \penalt_\Pi(M,l)$  and $(ii)$
for all integers $k > l$, $\penalt_\Pi(M',k)$ $=$ $\penalt_\Pi(M,k)$.
An answer set of $\Pi$ is \textit{optimal} if it is not dominated
by another one of $\Pi$. 
We also mention aggregates, an extension of ASP that we do not recall here for keeping simple the description. We refer the reader to ~\cite{aspcore2} for more details.

\begin{example}
Consider the following set of facts $F=\{\mathit{node}(1);$ $ \mathit{node}(2);$ $\mathit{node}(3);$ $\mathit{edge}(1,2);$ $\mathit{edge}(1,3);$ $\mathit{edge}(2,3)\}$, and the
ASP program $P$:
\begin{center}
$\begin{array}{l}
\mathit{col}(X,\mathit{red}) \vee \mathit{col}(X,\mathit{blue}) \vee \mathit{col}(X,\mathit{green}) \leftarrow \mathit{node}(X); \\  \leftarrow \mathit{edge}(X,Y),\ \mathit{col}(X,C), \ \mathit{col}(Y,C)
\end{array}$
\end{center} 
The set of facts $F$ models a cycle of length $3$, while the two rules of the program $P$ model the $3$-colorability problem. 
It can be checked,  that $F$ $\cup$ $\{\mathit{col}(1,\mathit{red})$, $\mathit{col}(2,\mathit{blue})$,
$\mathit{col}(3,\mathit{green})\}$ is an answer set of $P\cup F$.

To illustrate the usage of weak constraints, we now add the following 
$$\begin{array}{l}
\\ :\sim \ \naf \mathit{col}(X,\mathit{red}), \ \mathit{preferablyRed}(X). [1@1]
\end{array}$$
\noindent that prefers solutions having the nodes in $\mathit{preferablyRed}$ to be colored in red (note that each violation of the weak constraint increases the solution penalty by 1). Suppose we add to facts in input $\mathit{preferablyRed}(2)$, we obtain that an optimal solution is $F$ $\cup$ $\{\mathit{col}(1,\mathit{green})$, $\mathit{col}(2,\mathit{red})$, $\mathit{col}(3,\mathit{blue})\}$.
\end{example}
\color{black}

\begin{table*}[b] \small 
\begin{center}
    \caption{Base constructs of the annotation language.}\label{table:basic-annotations}
        \begin{tabular}{p{6.2cm} p{6.2cm}} 
        \hline
        \hline
        Annotation &  Description \\
        \hline\hline
        \texttt{@rule(name="\textit{rName}",block="\textit{bName}")} & The name \textit{rName} is assigned to the following rule. Assigning a rule to a block is optional.  \\ 
        \hline
        \texttt{@block(name="\textit{bName}",rules=\{\textit{rList}\})} & Defines a block with name \textit{bName}. Optionally a block may specify the list of rules that it covers.  \\
        \hline
        \texttt{@test( \newline
        \phantom{a}name = "\textit{testName}", \newline
        \phantom{a}scope = \{ \textit{referenceList} \}, \newline
        \phantom{a}programFiles = \{ \textit{programFileList} \}, \newline
        \phantom{a}input = "\textit{aspCode}", \newline
        \phantom{a}inputFiles = \{ \textit{inputFileList} \}, \newline
        \phantom{a}assert = \{ \textit{assertionList} \} \newline
        )} & Defines a test case with name \textit{testName} and scope \textit{referenceList} which is a list of strings referencing the rules and/or blocks under test. The target file is the current file, if \textit{programFiles} is not defined. An input for the test can be specified in \textit{aspCode} or several files (property \textit{inputFiles}) can be set optionally. Furthermore \textit{assertionList} is a list of assertions (defined in Table~\ref{table:assertions}) that have to be fulfilled for this test case.\\
        \hline
        \hline
        \end{tabular}
\end{center}
    \end{table*}

    \begin{table*}[h]\small 
    \caption{Assertions for \texttt{@test(...)} annotation.}\label{table:assertions}
        \begin{tabular}{p{7.5cm} p{4.8cm}} 
        \hline
        \hline
        Assertion &  Description \\
        \hline\hline
        \texttt{@noAnswerSet} & The test must have no answer set.  \\ 
        \hline
        \texttt{@trueInAll(atoms="\textit{atoms}")} & The atoms specified in \textit{atoms} must be true in all answer sets.  \\
        \hline
        \texttt{@trueInAtLeast(number=\textit{n},atoms="\textit{atoms}")} %
          & The atoms specified in \textit{atoms} must be true in at least \textit{n} answer sets. \\ 
        \hline
        \texttt{@trueInAtMost(number=\textit{n},atoms="\textit{atoms}")} & The atoms specified in \textit{atoms} must be true in at most \textit{n} answer sets.  \\ 
        \hline
        \texttt{@trueInExactly(number=\textit{n},atoms="\textit{atoms}")} & The atoms specified in \textit{atoms} must be true in exactly \textit{n} answer sets.  \\ 
        \hline
        \texttt{@constraintForAll(constraint="\textit{c}")} & The constraint specified in \textit{c} must be fulfilled in all answer sets.  \\
        \hline
        \texttt{@constraintInAtLeast(number=\textit{n},constraint="\textit{c}")} & The constraint specified in \textit{c} must be fulfilled in at least \textit{n} answer sets. \\ 
        \hline
        \texttt{@constraintInAtMost(}\texttt{number=\textit{n},constraint="\textit{c}")} & The constraint specified in \textit{c} must be fulfilled in at most \textit{n} answer sets. \\ 
        \hline
        \texttt{@constraintInExactly(}\texttt{number=\textit{n},constraint="\textit{c}")} & The constraint specified in \textit{c} must be fulfilled in exactly \textit{n} answer sets. \\ 
        \hline
        \texttt{@bestModelCost(cost=\textit{cv},level=\textit{lv})} & The best model has to meet the cost of \textit{cv} at level \textit{lv} (for weak constraints).  \\ 
        \hline
        \hline
        \end{tabular}
    \end{table*}

 \section{Unit Testing of Answer Set Programs}\label{sec:unit}

%
%
%


We now describe a new annotation-based test specification language that follows the Java annotation style of JUnit, and can be fully embedded in programs compliant with the ASP-Core-2 input language format of ASP competitions~\cite{DBLP:journals/jair/GebserMR17}, which is nowadays a common syntactic fragment supported by the main ASP implementations.
An annotation starts with '@' and is enclosed between \texttt{\%**} and \texttt{**\%} to distinguish multi-line comments, thus avoiding interference with program execution and to not require a separate test definition file (although in principle one could also collect testcases in separate files). 
The test specification language consists of base annotations and assertion condition annotations (or simply assertion annotations). The base annotations, described in Table~\ref{table:basic-annotations}, allow one to compose test cases, group subprograms in blocks, label rules and subprograms, and refer to the content of files containing programs. 
These annotations can be written anywhere in the ASP program, except \texttt{@rule(...)}, which has to be followed by an ASP rule in order to be assigned correctly. With regards to the \texttt{@test(...)} annotation the property \textit{scope} includes a list of strings as a parameter  that reference both rules and blocks under test (by their name). Furthermore the property \textit{assert} holds a list of assertion annotations that are described in Table~\ref{table:assertions}.
Basically, the programmer is free to identify the (sub)programs to test, specify the input of a program in a test case, and assert a number of conditions on the expected output, i.e., the basic operations supported by a XUnit testing language~\cite{Beck2002}.
Note that, we have considered in our proposal all the assertions that are both present in the main unit testing languages proposed in the literature and that are more frequent in our experience.

\begin{figure}[h]\small

\begin{lstlisting}[]
%** @block(name="ToTest") **%

%** Test graph **%
node(1). node(2). node(3). 
edge(1,2). edge(1,3). edge(2,3).

%** @rule(name="r1", block="ToTest") **%
col(X,red) | col(X,blue) | col(X,green) :- node(X).

%** @rule(name="r2", block="ToTest") **%
:- edge(X, Y), col(X,C), col(Y,C).

%**@test(name = "checkRules",
        scope = { "ToTest" },
        input = "node(1). node(2). node(3). edge(1,2). edge(1,3). edge(2,3).",
        assert = { 
        @trueInExactly(number = 2, atoms = "col(1, red)."),
        @trueInExactly(number = 1, atoms = "col(1, red). col(2, blue)") }
    )
**%
\end{lstlisting}
\caption{Testing graph colouring.}\label{lst:asp-3-color-problem} 
\end{figure} 
\begin{figure}[t] \small 
\begin{lstlisting}[]
%** @block(name="hamCycle") **%

%** @rule(name="r1", block="hamCycle") **%
inCycle(X,Y) | outCycle(X,Y) :- arc(X,Y).

%** @rule(name="r2", block="hamCycle") **%
reached(X) :- start(X).

%** @rule(name="r3", block="hamCycle") **%
reached(Y) :- reached(X), inCycle(X,Y).

%** @rule(name="r4", block="hamCycle") **%
:- inCycle(X,Y), inCycle(X,Z), Y<>Z.

%** @rule(name="r5", block="hamCycle") **%
:- inCycle(X,Y), inCycle(Z,Y), X<>Z.

%** @rule(name="r6", block="hamCycle") **%
:- node(X), not reached(X).

%** @test(name = "checkProperty",
         scope = { "hamCycle" },
         input = "node(1). node(2). node(3). node(4). arc(1,2). arc(1,4). arc(2,4). arc(3,1). arc(4,3). start(1)."
         assert = { @constraintForAll(":-node(X), #count{Y:inCycle(X,Y)}=0.") }
     )
**%
\end{lstlisting}
\caption{\reviewfix{Bugged encoding of Hamiltonian Cycle.}}\label{lst:asp-hamiltonian-path}
\end{figure}    

An usage example of the test annotations language can be found in Figure~\ref{lst:asp-3-color-problem}, which contains an instance of the graph coloring problem (3-colorability). This instance produces six answer sets according to the color assignments of the colors to the specified nodes. In order to test whether the rules behave as expected, we have to be able to reference the rules under test. As we do not want to test facts, we assign the names \textit{r1} and \textit{r2} to the rules in Lines 8 and 11. Additionally we assign these rules to a block, which has been defined in Line 1. Afterwards we are able to reference the rules under test inside the \texttt{@test(...)} annotation starting in Line 13. First we specify the name of the test case and the rules under test, which is the block \textit{ToTest} in this case. While referencing the block is more convenient, we could also reference the rules directly by writing \texttt{scope = \{ "r1", "r2" \}}. Input rules can be defined with the property \textit{input}, which are joined with the rules under test during test execution. They are equivalent to the facts of the program in this case, but can be different for more complex test specifications.  With the property \textit{assert} we can now define assertions that have to be fulfilled in order to execute the test with positive result. For this simple instance of the graph coloring problem, we can test whether the atom \texttt{col(1, red)} is true in exactly two answer sets while the atoms \texttt{col(1, red)} in combination with \texttt{col(2, blue)} should be true in exactly one answer set (Lines 17 and 18).

Note that, the \texttt{@test(...)} annotation is very flexible, and allows inputs  to be selected freely by picking any subprogram, which plays the role of a \textit{unit} to be tested (and run) in isolation. 
The \textit{scope} attribute can be filled with any list of references (cfr. Table~\ref{table:basic-annotations}), including single rules, lists of rules (mentioned by name), and rules conveniently collected in a block (as in the example). Thus, it is possible to fine tune tests selecting any subprogram the programmer wants to test. 
The programmer can also flexibly control the input, inserting specific facts, subprograms, or reading the additional inputs from a file.
In spite of being a simple ASP program, Figure~\ref{lst:asp-3-color-problem} shows how our lightweight annotation language can be used to define test cases without the need for a separate test definition file. The annotations do not interfere with program executability as being part of comments according to ASP-Core-2.

To further highlight the helpfulness of the TDD process for an ASP user, consider the example reported in Figure~\ref{lst:asp-hamiltonian-path}.
The program is a (buggy) ASP encoding for the Hamiltonian Cycle (HC) problem, a classical problem in graph theory. 
Given a finite directed graph $G = (N,A)$, and a node $a \in N$, the HC problem asks whether a cycle in $G$ exists starting from $a$ and passing through each node in $G$.
In our encoding, the first rule represents a guess for the set of arcs of the graph.
The second and the third rule model the reachability in a graph. Indeed, the starting node $X$ is reached (rule r2, line 7), and if $X$ is reached and $(X,Y)$ is in the cycle, then Y is also reached (rule r3, line 10).
Finally, the last three rules are constraints to be satisfied so that the arcs chosen in the cycle form a Hamiltonian cycle.
\reviewfix{
Indeed, the first two constraints state that for each node there is no more than one outgoing arc (rule r4, line 13) and there is no more than one incoming arc (rule r5, line 16); and the last constraint states that there is no node $X$ which is not reached (rule r6, line 19). 
}
Now, if we have found a Hamiltonian cycle, we expect to see in each answer set an outgoing arc for each node appearing in the cycle.
We can express this condition through a constraint, stating that it is not possible that we have a node $X$, and there is no arc from $X$ to some other node in the cycle. 
This condition is modeled by the assertion 
\texttt{@constraintForAll} in line 24, by meaning that each solution (answer set) must satisfy that condition.
If we consider the input
$A=\{node(1),$
$node(2),$
$node(3),$
$node(4),$
$arc(1,2),$
$arc(1,4),$
$arc(2,4),$
$arc(3,1),$
$arc(4,3),$
$start(1)
\}$ 
reported in lines 23-24, we expect that a Hamiltonian cycle exists. 
Note that, any programmer would run such kind of ``live" test, and maybe more than one, as in any programming language.
However, the testing process fails.
Indeed, our program on the given input admits the answer set:
$A$ $\cup$ $\{inCycle(1,2),$ $inCycle(2,4),$ $inCycle(4,3),$ $outCycle(1,4),$ $outCycle(3,1)\}$, which does not satisfy the assertion. \reviewfix{(Actually, the program would be a correct encoding for a Hamiltonian Path).} 
Note that, if you are the author of a piece of code, you are less likely to see a mistake without ``trying" the program (you are ``expecting" your statements to be correct), and common practice is to run a small instance to see if the result is as we expect.
Thus, without an automated testing procedure one should check manually all the answer sets or resort to a script. 
Automatic testing makes this phase of the development easier and declarative.
Note that, since unit tests remain in the source code, once the program is updated, they are not lost (as it happens to manually-handled result-checking sessions). Tests can be run again gaining all the advantages of \textit{regression} testing~\cite{Beck2002}. If a test fails, bugs can be identified with a \textit{debugger}~\cite{DBLP:journals/tplp/BusoniuOPST13}.

More examples are reported in~\cite{tobias}.

%
%
%
%
%
%
%

\section{Computational Complexity}\label{sec:complexity}
\reviewfix{In this section, we first overview the complexity classes that will be mentioned in the paper; then, we study the computational complexity of the main assertion-checking tasks that one can specify with the specification language described in the previous section.}

\color{black}

\subsection{Preliminaires on complexity classes}
We recall some basic definitions that will be useful in the remainder of the paper. Hereafter, we assume the reader has basic knowledge of computational complexity~\cite{DBLP:books/daglib/0018514} and focus on the counting complexity classes.
Roughly, the counting problems address the issue of computing the number of solutions of an instance of a given problem. For these problems, a number of specific complexity classes were introduced \cite{DBLP:journals/siamcomp/Valiant79,DBLP:journals/sigact/HemaspaandraV95}.
In particular, the class of counting associated with problems in NP is denoted by \#P. More formally, according to \citeN{DBLP:journals/siamcomp/Valiant79}, the class \#P denotes the set of functions counting the number of accepting paths of NP machines.
This idea was generalized to arbitrary complexity classes by resorting to the following definition:

\begin{definition}[\citeN{DBLP:journals/sigact/HemaspaandraV95}]\label{def:sharpdotc}
For a standard complexity class $\mathcal{C}$, $\#\cdot\mathcal{C}$ denotes the set of functions such that for some $\mathcal{C}$-computable binary predicate $R$ and a polynomial $p$ it holds that for every input string $x$:
$$f(x) = \lVert \{y \mid p(\mid x \mid) = \mid y \mid \wedge\ R(x,y)\} \rVert$$
\end{definition}
Intuitively, each function $f \in \# \cdot \mathcal{C}$ counts the strings of polynomial length, denoted by $y$, with respect to the input size $\mid x \mid$, such that the predicate $R(x,y)$ holds.
It is worth recalling that, as it has been noted by \citeANP{DBLP:journals/sigact/HemaspaandraV95} \citeyear{DBLP:journals/sigact/HemaspaandraV95}, the class $\#P = \#\cdot P$.

Another useful definition will allow us to consider the cases in which we are interested in verifying lower/upper bounds to the number of solutions. 

\begin{definition}[\citeN{DBLP:journals/sigact/HemaspaandraV95}]\label{def:cdotc}
Let $\mathcal{C}$ be a standard complexity class, then $A \in C \cdot \mathcal{C}$ if there exists a function $f \in \#\cdot \mathcal{C}$ and a polynomial-computable function $g$ such that $x \in A \leftrightarrow f(x)\geq g(x)$. 
\end{definition}

Starting from Definition~\ref{def:cdotc}, it can be verified that the complexity class $PP = C \cdot P$~\cite{DBLP:journals/sigact/HemaspaandraV95}.
\begin{table}[t]
\color{black}

    \centering
    \begin{tabular}{|r|c|c|c|}
        \cline{1-4}
        ASP Program     & Answer Set Checking  & Answer Set Existence  & Answer Set Counting \\
        \cline{1-4}
        Head-Cycle Free & P            & NP-c          & \#P-c \\
        \cline{1-4}
        Disjunctive     & coNP-c       & $\Sigma_2^P$-c  & \# $\cdot$ coNP-c \\ 
        \cline{1-4}
    \end{tabular}
    \caption{Complexity results of the main tasks concerning ASP programs evaluation.}
    \label{tab:complexityRecall}
\end{table}
Concerning the application of these definitions to the realm of ASP, 
the complexity of the Answer Set Counting problem has been extensively studied over the years~\cite{DBLP:journals/corr/FichteHMW16}.
In Table~\ref{tab:complexityRecall} we recall the complexity of the main tasks associated with the problem of checking, computing, and counting answer sets of ASP programs.
The results on the complexity of decision problems are summarized in a survey by \citeN{DBLP:journals/csur/DantsinEGV01} and derived in previous papers by \citeN{DBLP:journals/jacm/MarekT91} for normal programs, \citeN{DBLP:journals/amai/EiterG95} for disjunctive programs, and \citeN{DBLP:journals/amai/Ben-EliyahuD94} for the Head-Cycle Free (HCF) programs. The results on counting were provided by \citeN{DBLP:journals/corr/FichteHMW16}.

\subsection{Complexity of testing tasks}
For evaluating the complexity of each assertion task we considered the case for propositional ASP programs. In particular, Table~\ref{tab:Complexity} reports the corresponding complexity class for each assertion task.

\color{black}


\begin{theorem}
All results stated in Table~\ref{tab:Complexity} do hold.
\end{theorem}

In the following, we provide detailed proofs of the computational complexity of the implemented assertion tasks.
Note that, some of them can be easily obtained as corollary of well-known results about standard reasoning tasks in ASP, such as \textit{answer set existence}, \textit{brave reasoning}, and \textit{cautious reasoning}~\cite{DBLP:journals/csur/DantsinEGV01}, but, to the best of our knowledge, many others, in particular tasks that we identify to be $\mathsf{PP}$-complete and $\mathsf{C}\cdot\mathsf{coNP}$-complete, are new.   

\begin{table}[t]
  \begin{center}
    \begin{tabular}{lcc} 
   \hline \hline 
    \textbf{Assertions} & \hspace*{0.5cm}\textbf{Head-Cycle Free} \hspace*{0.5cm} & \ \textbf{Disjunctive} \\
   \hline \hline
    \texttt{@noAnswerSet} & $\mathsf{coNP}$-c& $\Pi_2^P$-c\\
    \hline 
	\texttt{@trueInAll(atoms)} & $\mathsf{coNP}$-c & $\Pi_2^P$-c\\
	\hline 
	\texttt{@trueInAtLeast(atoms,k)} & $\mathsf{PP}$-c  & $\mathsf{C}\cdot\mathsf{coNP}$-c\\
	\hline 
	\texttt{@trueInAtMost(atoms,k)} & $\mathsf{PP}$-c &$\mathsf{C}\cdot\mathsf{coNP}$-c\\
	\hline 
	\texttt{@trueInExactly(atoms,k)} & $\mathsf{PP}$-c &$\mathsf{C}\cdot\mathsf{coNP}$-c\\
	\hline 
	\texttt{@constraintForAll(constr)} & $\mathsf{coNP}$-c &$\Pi_2^P$-c\\
	\hline 
	\texttt{@constraintInAtLeast(constr,k)} & $\mathsf{PP}$-c &$\mathsf{C}\cdot\mathsf{coNP} $-c\\
	\hline 
	\texttt{@constraintInAtMost(constr,k)} & $\mathsf{PP}$-c &$\mathsf{C}\cdot\mathsf{coNP} $-c\\
	\hline 
	\texttt{@constraintInExactly(constr,k)} & $\mathsf{PP}$-c &$\mathsf{C}\cdot\mathsf{coNP}$-c\\
	\hline 
	\texttt{@bestModelCost(cost,level)}&$\Delta_2^P$-c & $\Delta_3^P$-c\\
   \hline \hline
    \end{tabular}
  \end{center}
  \caption{Computational complexity of the assertion tasks.}\label{tab:Complexity}
\end{table}


In the following, assume that the program $P$ is propositional.

%

\paragraph{\bf%
\texttt{@noAnswerSet}}
 asks whether $P$ has no answer set, i.e., $AS(P)=\emptyset$.
Hence, this problem is the complementary of the answer set existence problem.
Since answer set existence is $\mathsf{NP}$-complete for head-cycle free ASP programs and $\Sigma_2^P$-complete for disjunctive ASP programs~\cite{DBLP:journals/csur/DantsinEGV01}, then the assertion is $\mathsf{coNP}$-complete for head-cycle free programs, and $\Pi_2^P$-complete for disjunctive ASP programs.

\paragraph{\bf \texttt{@trueInAll}.}
Given a set of atoms $A$,
\texttt{@trueInAll($A$)} asks whether each atom in the set $A$ belongs to each answer set of $P$, i.e., for each $M\in AS(P)$, $A\subseteq M$? 
Hence, this assertion task is equivalent to the cautious reasoning task.
Since cautious reasoning is $\mathsf{coNP}$-complete for head-cycle free ASP programs and $\Pi_2^P$-complete for disjunctive ASP programs~\cite{DBLP:journals/csur/DantsinEGV01}, then the assertion is also $\mathsf{coNP}$-complete for head-cycle free programs, and $\Pi_2^P$-complete for disjunctive ASP programs.

\paragraph{\bf \texttt{@trueInAtLeast}.}
Given a set of atoms $A$ and a positive integer $k$,
\texttt{@trueInAtLeast($A,k$)} 
\reviewfix{asks whether there exist $M_1,\ldots,M_k \in AS(P)$ such that $A \subseteq M_i$, with $1\leq i \leq k$.
Note that, this problem is equivalent to check if, the ASP program $P\cup\{\leftarrow not \ a \mid a \in A\}$ has at least $k$ answer sets.} 
Now, it is known that the problem of answer set counting is $\#\mathsf{P}$-complete for head-cycle free programs, and is $\#\cdot\mathsf{coNP}$-complete for disjunctive programs~\cite{DBLP:journals/corr/FichteHMW16}. Hence, the corresponding decision problem of checking if the number of answer sets is at least $k$ is $\mathsf{PP}$-complete for head-cycle free programs, and is $\mathsf{C}\cdot\mathsf{coNP}$-complete for disjunctive programs.%
\footnote{For further details on complexity classes related to counting problems, we refer to~\cite{DBLP:journals/sigact/HemaspaandraV95}.}

\paragraph{\bf \texttt{@trueInAtMost}.}
Given a set of atoms $A$ and a positive integer $k$,
\texttt{@trueInAtMost($A,k$)} 
\reviewfix{asks whether there exist at most $k$ answer sets, $M_1,\ldots,M_k \in AS(P)$ such that $A \subseteq M_i$, with $1\leq i \leq k$.
Moreover, this problem is equivalent to check if, the ASP program $P\cup\{\leftarrow not \ a \mid a \in A\}$ has at most $k$ answer sets.}
It is known that, the decision problem of checking if the number of satisfying assignments for a given Boolean formula is at most $k$, has the same complexity of checking if the number of  satisfying assignments is at least $k$~\cite{DBLP:journals/siamcomp/Valiant79}.
Thus, the assertion task @trueInAtMost has the same computational complexity of @trueInAtLeast.

\paragraph{\bf \texttt{@trueInExactly}.}
Given a set of atoms $A$ and a positive integer $k$,
\texttt{@trueInExactly($A,k$)} 
\reviewfix{asks whether there exist exactly $k$ answer sets, $M_1,\ldots,M_k \in AS(P)$ such that $A \subseteq M_i$, with $1\leq i \leq k$.}
Note that the assertion \texttt{@trueInExactly($A,k$)} is true if, and only if, \texttt{@trueInAtLeast($A,k$)} and \texttt{@trueInAtMost($A,k$)} are true.
Therefore, the assertion task \texttt{@trueInExactly} has the same computational complexity of the assertion tasks \texttt{@trueInAtLeast} and \texttt{@trueInAtMost}.

\paragraph{\bf%
\texttt{@constraintForAll}.}
Given a set of constraints $C$,
\texttt{@constraintForAll} asks whether each answer set of $P$ is also an answer set of $P\cup C$, i.e., $AS(P)=AS(P\cup C)$.
This problem is equivalent to ask whenever $P\cup C'$ has no answer set, where $C' = \{fail \leftarrow B(c)\mid c\in C\} \cup \{ \leftarrow not \ fail \}$.
Indeed, assume that each answer set $M$ of $P$ is also an answer set of $P\cup C$. Hence, $M$ models each constraint in $C$. Therefore, $B(c)$ is never satisfied. Thus, $fail$ cannot be inferred, and it must be false. Hence, the constraint $\leftarrow not \ fail$ is not satisfied, and $P\cup C'$ has no answer set. The other implication is similar.
In conclusion, this problem is the complementary of the answer set existence problem~\cite{DBLP:journals/csur/DantsinEGV01}. Therefore, the assertion is $\mathsf{coNP}$-complete for head-cycle free programs, and $\Pi_2^P$-complete for disjunctive programs.
 
\paragraph{\bf%
\texttt{@constraintInAtLeast}.}
Given a set of constraints $C$ and a positive integer $k$,
\texttt{@constraint\-InAtLeast} asks whenever there exist at least $k$ answer sets of $P$ that satisfy the set of constraints $C$, i.e., if $P\cup C$ has at least $k$ answer sets.
Again, since the problem of answer set counting is $\#\mathsf{P}$-complete for head-cycle free programs, and is $\#\cdot\mathsf{coNP}$-complete for disjunctive programs~\cite{DBLP:journals/corr/FichteHMW16}, then, the problem of checking if the number of answer sets is at least $k$ is $\mathsf{PP}$-complete for head-cycle free programs, and $\mathsf{C}\cdot\mathsf{coNP}$-complete for disjunctive programs.

\paragraph{\bf%
\texttt{@constraintInAtMost}.}
Given a set of constraints $C$ and a positive integer $k$,
\texttt{@constraint\-InAtMost} asks whenever there exist at most $k$ answer sets of $P$ that satisfy the set of constraints $C$, i.e., if $P\cup C$ has at most $k$ answer sets. Note that this problem is equivalent to ask whenever $P\cup C$ has not at least $k+1$ answer sets, that is the complementary problem of \texttt{@constraintInAtLeast}, and it has the same complexity of \texttt{@constraintInAtLeast}.

\paragraph{\bf%
\texttt{@constraintInExactly}.}
Given a set of constraints $C$ and a positive integer $k$,
\texttt{@constraint\-InExactly} asks whenever there exist exactly $k$ answer sets of $P$ that satisfy the set of constraints $C$, i.e., if $P\cup C$ has exactly $k$ answer sets. Hence, this problem is equivalent to ask whenever $P\cup C$ has at least $k$ answer sets, and has at most $k$ answer sets.
Therefore, it has the same complexity of \texttt{@constraintInAtLeast} and \texttt{@constraintInAtMost}.

\paragraph{\bf \texttt{@bestModelCost}.}
Given a set of weak constraints $W$, and two positive integers $c$ and $l$, \texttt{@best\-ModelCost(c,l)} asks whenever there is an optimal answer set $M$ of $P\cup W$ such that its cost is $c$ at level $l$.
It is easy to see that the computational complexity of this task is equal to that of deciding the existence of an optimal answer set for programs with weak constraints, which is $\Delta_2^P$-complete for head-cycle free programs, and $\Delta_3^P$-complete for disjunctive programs~\cite{DBLP:journals/tkde/BuccafurriLR00}.


\begin{figure}[t]
\centering
\includegraphics[width=1\textwidth]{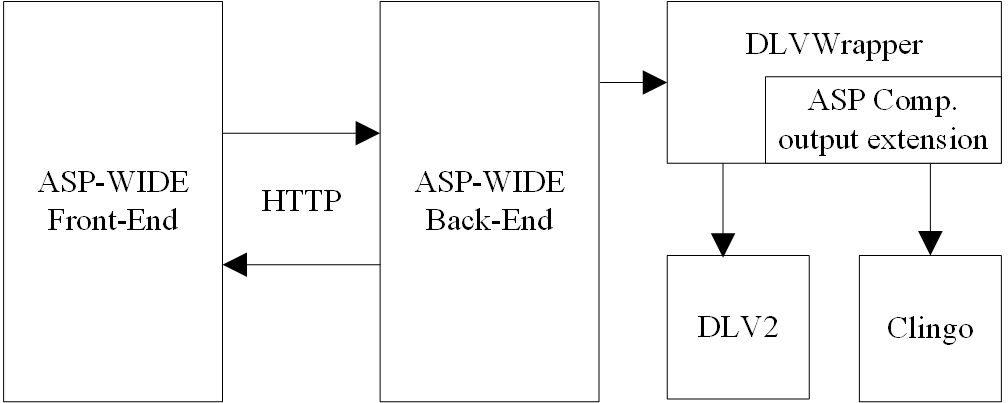}
\caption{Architecture of \reviewfix{ASP-WIDE}.}\label{fig:asp-wide-arc} 
\end{figure}

\section{The \reviewfix{ASP-WIDE} environment}\label{sec:env}
The \reviewfix{ASP-WIDE} environment implements this paper's unit-testing mechanism and the annotation language. While command line tools are efficient to use, the focus was to build an environment containing a code editor with syntax checking, syntax highlighting and execution/testing capabilities. This integrated development tool offers a convenient environment for writing, executing and testing answer set programs. Since web-based environments, not only for logic programming, but also for conventional languages, are widely used, \reviewfix{ASP-WIDE} is based mostly on web technologies. 

\subsection{Architecture and Implementation}
As many modern web-based applications, \reviewfix{ASP-WIDE} consists of a front-end, which is built using the Angular framework, and a back-end implemented in Java utilizing the Spring framework. 
The communication between front-end and back-end is realized with HTTP-Requests transmitting JSON data. 
The overall architecture is depicted in Figure~\ref{fig:asp-wide-arc}; below we detail the two main components of \reviewfix{ASP-WIDE}. \reviewfix{For a deeper (more technical) description of the implementation we refer the reader to the Master Thesis of Tobias Berei \cite{tobias}.}

\begin{table*}[t!] \small 
  \begin{center}

    \begin{tabular}{llp{5.2cm}} 
    \hline    \hline
	 \textbf{Assertions} &  \textbf{Tester Program} $T_P$ & \textbf{Test output} \\
    \hline    \hline
    \texttt{@noAnswerSet} & $P$ & Return $\mathit{fail}$ if $T_P$ admits answer sets, $\mathit{pass}$ otherwise.\\
    \hline
	\texttt{@trueInAll($A$)} & $P  \cup \bigcup_{a \in A} \{\leftarrow\ a\}$ & Return $\mathit{fail}$ if $T_P$ admits answer sets, $\mathit{pass}$ otherwise.\\
    \hline
	\texttt{@trueInAtLeast($A$,$k$)} & $P \cup \bigcup_{a \in A} \{\leftarrow\ \reviewfix{not\ a}\}$ & Return $\mathit{pass}$ as soon as the solver on $T_P$ outputs $k$ answer sets, $\mathit{fail}$ otherwise.\\
    \hline
	\texttt{@trueInAtMost($A$,$k$)} & $P \cup \bigcup_{a \in A} \{\leftarrow\ \reviewfix{not\ a}\}$ & Return $\mathit{pass}$ if the solver terminates on $T_P$ outputting at most $k$ answer sets, $\mathit{fail}$ otherwise.\\
    \hline
	\texttt{@trueInExactly($A$,$k$)} & $P \cup \bigcup_{a \in A} \{\leftarrow\ \reviewfix{not\ a}\}$ & Return $\mathit{pass}$ if the solver on $T_P$ outputs $k$ answer sets, $\mathit{fail}$ otherwise.\\
    \hline
	\texttt{@constraintForAll($C$)} & $P \cup \{f \leftarrow\ C;\ \leftarrow\ \reviewfix{not\ f}\}$ & Return $\mathit{pass}$ if $T_P$ admits no answer set, $\mathit{fail}$ otherwise.\\
    \hline
	\texttt{@constraintInAtLeast($C$,$k$)} & $P \cup \{C\}$ & Return $\mathit{pass}$ as soon as the solver on $T_P$ outputs $k$ answer sets, $\mathit{fail}$ otherwise.\\
    \hline
	\texttt{@constraintInAtMost($C$,$k$)} & $P \cup \{C\}$ &Return $\mathit{pass}$ if the solver terminates on $T_P$ outputting at most $k$ answer sets, $\mathit{fail}$ otherwise.\\
    \hline
	\texttt{@constraintInExactly($C$,$k$)} & $P \cup \{C\}$ &Return $\mathit{pass}$ if the solver on $T_P$ outputs $k$ answer sets, $\mathit{fail}$ otherwise.\\
    \hline
	\texttt{@bestModelCost(\textit{c,l})}& $P$ & Return $\mathit{pass}$ if the optimal answer set of $T_P$ has a cost of $c$ at level $l$, $\mathit{fail}$ otherwise.\\
    \hline    \hline
    \end{tabular}
    
  \end{center} \color{black}
  \caption{Implementation of the assertions. $P$ denotes the scope (i.e., the subprogram to test), $T_P$ the program built to implement the assertion, $C$ a constraint, $A$ a set of atoms, $k$, \textit{c}, \textit{l} are integers.}\label{tab:execution}
\end{table*}

\paragraph{The front-end} of the development environment was built with the Angular framework (see \url{https://angular.io/}), which is a web technology for building comprehensive web applications. The UI components are partly based on Angular Material components (cfr. \url{https://material.angular.io/}), while the code editor utilizes the Monaco editor 
known from Visual Studio Code (see \url{https://microsoft.github.io/monaco-editor/index.html}). 
The Monaco editor comes with several built-in features, like the possibility to define a custom syntax highlighting using Monarch,
but also syntax errors/warnings can be displayed efficiently. 
Thus \reviewfix{ASP-WIDE} environment is restricted to run inside a browser that renders HTML with CSS and executes JavaScript in background.


\paragraph{The back-end} of \reviewfix{ASP-WIDE} consists of a REST web service implemented in Java using the Spring framework. In particular, web controllers receive web requests and execute the required logic in order to fulfill actions for the front-end. Some of these actions are: $(i)$ file management operations; $(ii)$ checking the syntax of the program; $(iii)$ executing programs and returning answer sets; and $(iv)$ executing unit-tests and returning test results.
Following ``Separation of concerns'', a well-known design principle, certain aspects of the back-end have been split into three modules, namely \textit{Development}, \textit{Execution} and \textit{File Management}.
\begin{figure}
\begin{lstlisting}[]
public class TestSuite{
    private HashMap<String, Block> blocks;
    private HashMap<String, String> rules;
    private ArrayList<Test> tests;
    public TestSuite() {
        this.blocks = new HashMap<String, Block>();
        this.rules = new HashMap<String, String>();
        this.tests = new ArrayList<Test>();
    }
    // further functions and definitions
}
\end{lstlisting}
\caption{\texttt{TestSuite} class definition.}\label{lst:test-suite-class} 
\end{figure} 

\reviewfix{The File Management module deals with typical CRUD operations for files, while the Development and Execution modules implement the \emph{testing engine} that is responsible for syntax checking and execution of programs and tests. }

\reviewfix{The main steps of the proposed test engine can be summarized as follows:
(i) parsing of the ASP-Core-2 input language including an extension for the annotation language of this paper; (ii) interpreting parsed assertions in order to generate the tester program; (iii) execution of ASP systems; and (iv) evaluating test results.  
}

\reviewfix{More precisely, the test engine starts by parsing the entire ASP program, together with provided test annotations, by means of a parser generated with JavaCC (see \url{https://javacc.org/}). In particular, such a parser is obtained from a grammar definition of the input language and provides the possibility to implement a visitor pattern on top of the grammar tree by means of JJTree tool included in JavaCC. During parsing, exceptions are caught by our engine in order to report to the users all useful information (i.e., error line and column) to identify the error in the provided ASP code. If no errors happen at this stage, an object-oriented representation of the parsed program and the parsed annotations is obtained.}

\reviewfix{In particular, this is encapsulated in the Java class named \texttt{TestSuite} reported in Figure~\ref{lst:test-suite-class}. By exploiting such a virtual model, rules and blocks of the original program are stored only once and are shared among the different tests, while test-specific data such as assertions, input facts, constraints, and more, are enclosed in the different \texttt{Test} objects.}


\reviewfix{Starting from the obtained \texttt{TestSuite} each test is executed separately. In order to better understand the testing workflow, Figure~\ref{lst:trueInAtMost} reports the code that implements the execution of a  test verifying a \texttt{@trueInAtMost} assertion.}


\reviewfix{The function \texttt{checkAssertion} receives three parameters, an ASP program (\texttt{code}), an assertion to verify (\texttt{assert}), and the ASP system that should be used (\texttt{st}). Intuitively, the ASP program is obtained from the generated \texttt{TestSuite}, by fetching all the rules indicated (explicitly or indirectly by using block definitions) in the scope attribute of the test while the assertion object stores the answer set count $k$, and the list of $atoms$ that has to be true in at most $k$ answer sets.}

\begin{figure}
\begin{lstlisting}[]
    private AssertionResult checkAssertion(String code, AssertTrueInAtMost assert, SolverType st) {
        AssertionResult ar = new AssertionResult();
        ar.setName(assert.getClass().getSimpleName());
        ar.setExecutedCode(code);
        String[] atoms = assert.getAtoms().split("\\.");
        StringBuilder sb = new StringBuilder(code);
        for(String atom: atoms) {
            sb.append(":- not ");
            sb.append(atom.trim());
            sb.append(".\n");
        }
        ar.setExecutedCode(sb.toString());
        ExecutableFile testFile = new ExecutableFile("checkAssertionTrueAtMost", ar.getExecutedCode(), "");
        testFile.setSolverType(st);
        ExecutionResult er = this.executionLogic.executeCode(testFile, (assert.getAssertCount()+1));
        ar.setExecutionOutput(er.getResult());
        if(er.getModels() != null && er.getModels().size() <= assert.getAssertCount()) {
            ar.setSucceeded(true);
        }
        return ar;
    }
\end{lstlisting}
\caption{Implementation of the \texttt{@trueInAtMost} assertion.}\label{lst:trueInAtMost} 
\end{figure} 

\reviewfix{According to the transformation described in Table~\ref{tab:execution}, in our example, the tester program is obtained by adding to the input program a constraint for each atom that has to be true (Figure~\ref{lst:trueInAtMost}, lines 6-11).}

\reviewfix{Then, the tester program is evaluated by the ASP system \texttt{st} (Figure~\ref{lst:trueInAtMost}, lines 12-15). This evaluation is implemented in the Execution module that is built on top of the library DLVWrapper~\cite{DBLP:conf/asp/Ricca03}. In particular, we extended the DLVWrapper library to handle any ASP systems supporting the output format of the last ASP Competition~\cite{DBLP:journals/jair/GebserMR17}, such as e.g., Clingo~\cite{DBLP:journals/tplp/GebserKKS19}. According to the selected ASP system, DLVWrapper executes it as a command-line tool with its specific arguments. A snapshot of the code that prepares the execution engine is reported in Figure~\ref{lst:dlvwrapper}.}

\reviewfix{There are several benefits that come with the DLVWrapper. 
One of them is that it provides a layer of objects that abstracts the interaction with the solver and simplifies the implementation. Moreover, it supports asynchronous program execution with the possibility to register callback functions (ModelHandlers) for found answer sets, which allows handling large results without saturating the memory of the caller.}

\begin{figure}
\begin{lstlisting}[]
protected void startRun throws IOException, DLVInvocationException {
    
    // some preparation ...
    
    List<String> input = new ArrayList<String>();
    input.add(DLVWrapper.getInstance().getPath());
    try {
        if (isDLV2){
            addOption("--silent");
        }
        else if(isClingo) {
            addOption("--outf=1");      // competition output format
            addOption("--quiet=0,0");   // output configuration
        }
        else{ // DLV
            addOption("-silent");
        }
    } catch (DLVInvocationException e) {
        e.printStackTrace();
    }
    input.addAll(options);
    input.addAll(inputProgram.getFiles());
    
    // starting the ASP system in a process ...

}
\end{lstlisting}
\caption{Customization of DLVWrapper according to the used ASP system.}\label{lst:dlvwrapper} 
\end{figure} 
\begin{figure}[t]
\centering
\includegraphics[width=\textwidth]{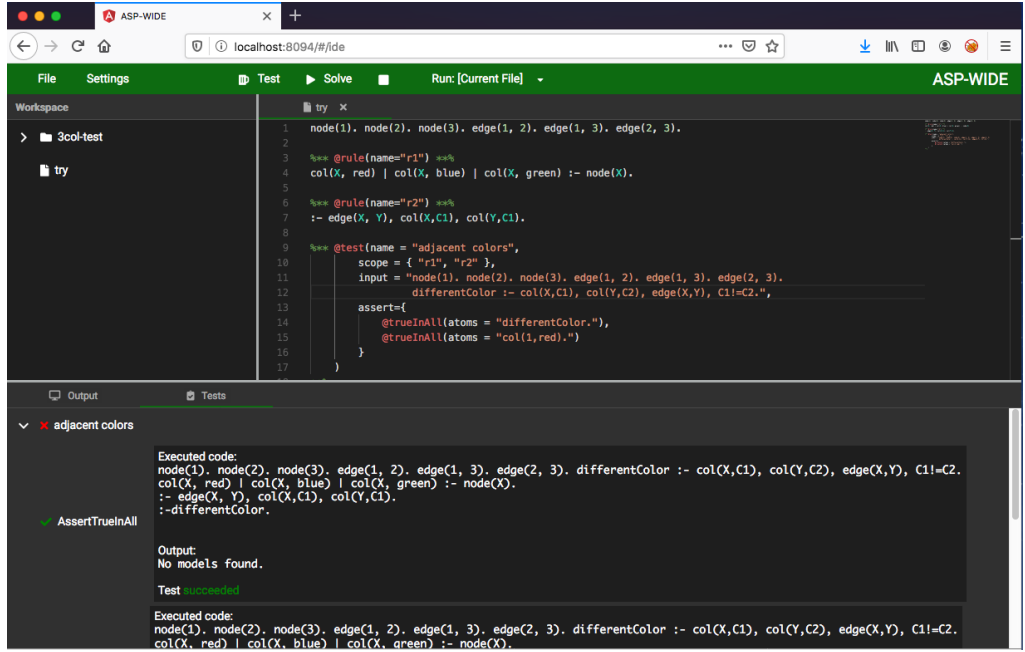}
\caption{The \reviewfix{ASP-WIDE} interface.}\label{fig:aspwide}
\end{figure}
\reviewfix{The output produced by the ASP system (i.e., the list of answer sets of the tester program) is encapsulated in an object of the class \texttt{ExecutionResult} that is further used to check the tested assertion. In our example, if at most $k$ answer sets have been found out of the $k+1$ requested by the execution module, then the assertion succeeded (Figure~\ref{lst:trueInAtMost} lines 16-19). Assertion results are then provided as output to the user request and are visualized by the front-end.}
\subsection{User interface} 
The user interface of \reviewfix{ASP-WIDE} is depicted in Figure~\ref{fig:aspwide}.
The design of the environment was inspired by ASPIDE and features four main areas of interaction with the user: 
\begin{itemize}
\item[$(i)$] The toolbar on the top of the environment;
\item[$(ii)$] The workspace or file explorer on the left side;
\item[$(iii)$] The code editor in the middle/right area (with open tabs on the top); and
\item[$(iv)$] The output area on the bottom, which shows answer sets and test results.
\end{itemize}

The toolbar features usual menus for handling files and settings; two buttons to run files and tests and a drop-down list for handling solver execution configurations (called run configuration as in ASPIDE), respectively.
Programs can be organized in projects, which appear and can be operated as folders on the workspace explorer.
Acting on the workspace explorer one can open files. The opened file is displayed in the code editor in the middle/right area of the screen. The code editor features syntax-highlights and code-completion, i.e., it suggest how to complete predicate names and variables, to assist program and test case development. Errors and warnings are also immediately displayed, and the modifications are automatically saved, as is customary in modern web-based interfaces.
The execution of programs can be controlled by managing run configurations, where one can specify the files to include, the solver to use and the command line parameters. 
Executing the current file or unit-tests (if there are any) is done by clicking on the buttons ``Solve'' or ``Test'', which can be found in the middle of the Toolbar. 
In case one just wants to run some files together,  the interface allows the user to select files from the workspace explorer, right click and select ``Run directly'' from a drop-down menu. 
More involved configurations can be set up acting on the run configuration window.
The result of the execution (of tests and programs) is shown in the bottom part of the environment, containing output area. The result of test case execution outlines, for each test, the result (using red color for failed tests, and green color for passed ones), and for each test it is possible to inspect the execution details, and a witness of the result.

\subsection{Availability}
The \reviewfix{ASP-WIDE} environment can be installed as a standalone application on any computer with a modern (java-script enabled) web browser and Java 8 installed. 
\reviewfix{ASP-WIDE}  can be downloaded 
from: 
\url{http://www.mat.unical.it/ricca/asptesting/asp-wide_1.0-beta4.zip}, 
and the sources are distributed under GPL license
(send an email to  \url{ricca@mat.unical.it}).

\newcommand{\mutation}{\textsc{ASP-Modificator}\xspace}

\section{Experimental Evaluation}\label{sec:experiments}
In this section we present an experimental evaluation aimed at assessing the \reviewfix{ASP-WIDE} implementation of the test case execution environment. 
On one hand, the experiment aims to demonstrate the practical usability of the \reviewfix{ASP-WIDE} language by showcasing its ability to effectively identify and capture bugs in real-world use cases. 
Also, the experiment shows that \reviewfix{ASP-WIDE} exhibits good performance in test case execution.
On the other hand, the experiment highlights the importance of \reviewfix{having} a complexity-aware implementation of test case execution.

\subsection{Development of the test suites}
The first step of the experimental campaign was the development of a \reviewfix{benchmark covering} some concrete use case problems. 
In particular, we considered three well-known problems used in the literature both to introduce ASP and to test ASP systems in competitions~\cite{DBLP:journals/tplp/GebserMR20,DBLP:journals/jair/GebserMR17,DBLP:series/synthesis/2012Gebser}: (HP) Hamiltonian Path, (FF) Fast Food problem, and Quantified Boolean Formulas with two quantifiers (2QBF). 

\reviewfix{The HP problem involves determining the existence of a finite set of vertices $v_1,\ldots,v_k$ of a directed graph $G=\langle V,E \rangle$, such that: (i) every node $v \in V$ appears exactly once, and (ii) for each $1\leq i < k$, $(v_i,v_{i+1}) \in E$.}
HP is a well-known problem in literature belonging to the class of NP-complete problems. 
HP played a relevant role in the ASP literature because it is one of the canonical examples of non-tight~\cite{DBLP:journals/tplp/ErdemL03} encoding (i.e., with positive recursive definitions).

FF is an optimization problem introduced in ASP competitions~\cite{DBLP:journals/tplp/GebserMR20}.
The objective of FF is to find \reviewfix{a $k$-subset} of $n$ restaurants that serve as depots in such a way the distance between each restaurant and the closest depot is minimized. 
FF features an encoding that follows the guess\&check methodology and makes use of aggregates and, importantly, of weak constraints to model the optimality condition.

2QBF is the problem of deciding the satifiability of a quantified boolean formula with two quantifiers.
More in detail, the problem is to check satisfiability of a formula of the form $\exists X\forall Y\phi$, where $X$ and $Y$ are sets of propositional variables, and  $\phi$ is a formula in disjunctive normal form. 
Recall that a formula $\phi$ is in disjunctive normal form if it is of the form $C_1 \vee \cdots \vee C_n$ where $C_i$ is a conjunction of propositional literals (i.e. propositional variables or their negation) for all $i=1,\cdots,n$. 2QBF is $\Sigma_2^P$-complete.
Note that, 2QBF is the canonical example of the application of the saturation programming technique~\cite{DBLP:journals/amai/EiterG95}, that is used to model problems belonging to the second level of the \reviewfix{Polynomial Hierarchy} (PH). 

All in all, the experiment covers many of the main constructs of ASP (i.e., recursive definitions, constraints, aggregates) as well as the two main programming methodologies (i.e., guess\&check and saturation). Moreover, it considers representatives for both decision and optimization problems belonging to different levels of the PH.
The encodings of HP, FF and 2QBF have been introduced in the literature and used also in ASP competitions~\cite{DBLP:journals/tplp/GebserMR20}.
Thus, this benchmark can be pragmatically considered to be made of concrete and representative use cases of ASP programs.

\begin{table*}[t]
    \centering
    \begin{tabular}{@{\extracolsep{\fill}} lccc}
    \hline\hline
    Assertion$\backslash$Domain  & Hamiltonian Path  & Fast Food     & 2QBF \\
    \hline\hline
    \texttt{@constraintForAll}          & $\checkmark$      & $\checkmark$  & $\checkmark$\\
    \texttt{@constraintInAtLeast}       & -                 & -             & $\checkmark$\\
    \texttt{@trueInAll}                 & -                 & -             & $\checkmark$\\
    \texttt{@trueInAtLeast}             & -                 & $\checkmark$  & $\checkmark$\\
    \texttt{@trueInExactly}             & $\checkmark$      & -             & -\\
    \texttt{@noAnswerSet}               & $\checkmark$      & -             & $\checkmark$\\
    \texttt{@bestModelCost}             & -                 & $\checkmark$  & -\\
    \hline\hline
    \end{tabular} 
    \caption{Employed assertions for each domain. \label{tab:used_assertion}}
\end{table*}

A test suite for the above-mentioned problems has been developed by using the testing language presented in Section~\ref{sec:unit}.
We have created unit test specifications for the considered problems: HP, FF, and 2QBF, consisting of 6, 5, and 6 cases, respectively.
Table~\ref{tab:used_assertion} summarizes the assertions used in the test cases for each considered problem. 
We observe that \texttt{@constraintForAll} assertion, checking properties that hold in all answer sets, was used in all the three cases; whereas \texttt{@trueInExacly} revealed to be useful only in HP. As expected, \texttt{@bestModelCost} was used only in FF, which is the only optimization problem. 
All the remaining assertions were used in at least two cases.
The test case specifications were accompanied by random problem instances, forming a test suite consisting of 88 unit tests for HP, 23 for FF, and 459 for 2QBF.

\begin{table*}[b]
    \caption{Modifications applied to generate mutants. \label{tab:used_modification}}
    \centering
    \begin{tabular}{@{\extracolsep{\fill}} lccc}
    \hline\hline
    Modification$\backslash$Domain  & Hamiltonian Path  & Fast Food     & 2QBF \\
    \hline\hline
    renamePredicates        & $\checkmark$ & $\checkmark$ & $\checkmark$\\
    deleteRule              & $\checkmark$ & $\checkmark$ & $\checkmark$\\
    deleteLiteral           & -            & $\checkmark$ & $\checkmark$\\
    addDefaultNegation      & $\checkmark$ & -            & $\checkmark$\\
    swapTerms               & $\checkmark$ & -            & $\checkmark$\\
    changeAggregates        & -            & $\checkmark$ & -\\
    changeMathOperators     & -            & $\checkmark$ & -\\
    swapDefaultNegation     & -            & $\checkmark$ & -\\
    \hline\hline
    \end{tabular} 
\end{table*}

In order to assess our test suite, we performed a mutation analysis~\cite{DBLP:conf/kr/OetschPPST12}.
\reviewfix{The latter} involves making various modifications to the \textit{original} (correct) encoding, resulting in the creation of \textit{mutants}. 
The purpose is to evaluate the effectiveness of the test suite in detecting potential bugs by executing it on these mutants.
More in detail, we utilized the mutation model and tool specifically designed for ASP, which was introduced by \nCite{DBLP:conf/kr/OetschPPST12}. 
In our work, we refer to this tool as \mutation\reviewfix{, which} provides a pool of possible modifications that can be applied, possibly multiple times, to an input program $P$, in order to generate different \textit{mutants} of $P$.
Among possible modifications supported by \mutation, we selected those \reviewfix{that can be applied to the considered} problem encoding according to their syntactic features. 
In particular, Table~\ref{tab:used_modification} reports the list of modification used for generating mutants for each problem. 
By applying such transformations, we obtained 9 altered mutants (i.e. $M_1,\cdots,M_9$) for each problem encoding.

\subsection{Experiment setup}
In order to ensure the reproducibility of our results, we provide detailed information below regarding the methods compared and the execution environment used in the study.

\paragraph{Compared methods.}
In the experiment we run two implementations.
The first is the test case execution engine of \reviewfix{ASP-WIDE}, configured with the ASP system \textsc{clingo}~\cite{DBLP:journals/tplp/GebserKKS19}.
Recall that, \reviewfix{ASP-WIDE} produces for each test case a tester program (as specified in Table~\ref{tab:execution}), and then analyses the output of the ASP solver on the tester program to present the results to the user. This behavior has been extracted from the environment and incorporated into a command line tool, which is then executed in isolation within the experimental environment.

Another implementation that was executed is a re-engineering of the algorithms employed in ASPIDE, which we denote as \reviewfix{ASPIDE-LIKE}.
Note that, ASPIDE does not provide direct access to the internal engine, thus it has been faithfully re-implemented.
More in detail, in the ASPIDE approach the semantics of the assertions is implemented straightly. In few words, the instance under test is assembled according to the test case specification, and fed to an ASP system (i.e., \textsc{clingo} in this experiment). In turn, a Java thread processes the answer sets to check whether the assertion to test is satisfied. 
This approach is not complexity-wise optimal in most cases.
For example, checking \reviewfix{whether} a property holds in all answer sets requires a full enumeration in ASPIDE, whereas it can be done in a single call to an ASP solver in \reviewfix{ASP-WIDE} (cfr. Table~\ref{tab:execution}).
\reviewfix{ASPIDE-LIKE} implementation is ``complexity-wise optimal'' only when analyzing the first solution generated by the ASP system is adequate, e.g., in the case of \texttt{@bestModelCost}.

\paragraph{Hardware and software resources.}
All the experiments were executed on a machine equipped with Xeon(R) Gold 5118 CPUs, running Ubuntu Linux (kernel 5.4.0-77-generic). Memory and time were limited to $4$ gigabyte and $300$ seconds respectively. 
The evaluation of ASP programs was performed by running \textsc{clingo}~\cite{DBLP:journals/tplp/GebserKKS19} version 5.4.0.

\paragraph{Benchmarks availability.}
All the material (encoding, instances, tests, executables, etc.) needed to reproduce the experiments can be downloaded from \url{https://osf.io/6hxdn/?view_only=2a2067f712ac438cb3b6a9c7aa528331}.
\newcommand{\imagesSize}{0.78}
\usetikzlibrary{
    patterns,
}

\begin{figure*}[ht!]

    \begin{subfigure}[t]{0.8\textwidth}
        \centering
        \vspace*{0.5cm}
        \includegraphics[width=0.58\textwidth]{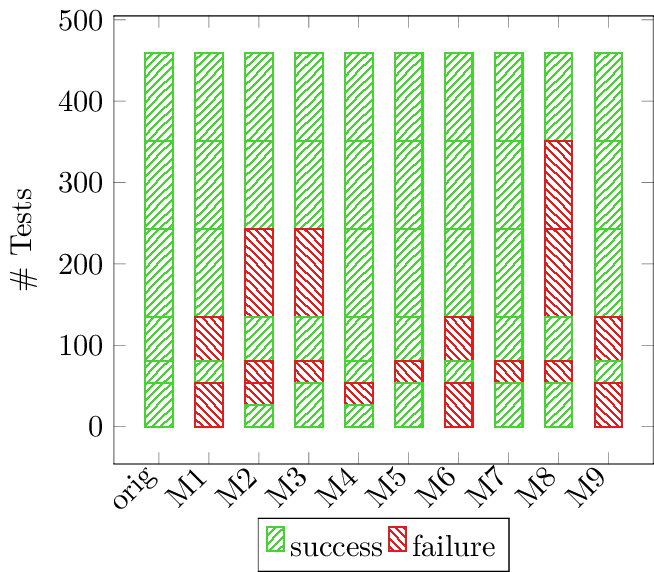}
        \caption{Unit tests (2QBF)\label{fig:unit_2qbf}}
    \end{subfigure}
    \begin{subfigure}[t]{0.8\textwidth}
        \centering
        \vspace*{0.5cm}
        \includegraphics[width=0.58\textwidth]{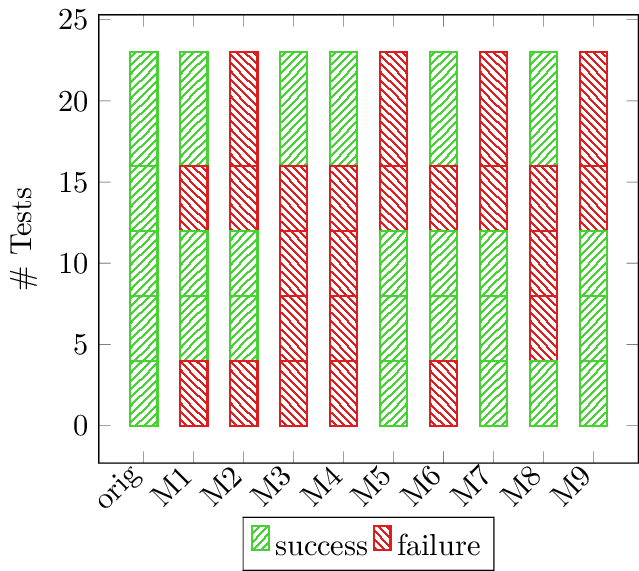}
        \caption{Unit tests (FF)\label{fig:unit_ff}}
    \end{subfigure}
    \begin{subfigure}[t]{0.8\textwidth}
        \centering
        \vspace*{0.5cm}
        \includegraphics[width=0.58\textwidth]{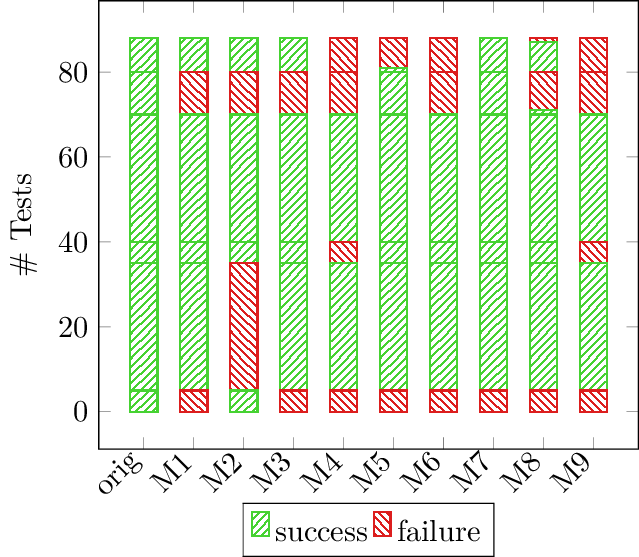}
        \caption{Unit tests (HP)\label{fig:unit_hp}}
    \end{subfigure}
    \caption{Successful and failed unit tests for each domain.\label{fig:bar_plot}}
\end{figure*}

\subsection{Results}

In the following the results obtained by running \reviewfix{ASP-WIDE} and \reviewfix{ASPIDE-LIKE} are analyzed.

\paragraph{Efficacy of the testing methodology.}
First we discuss about the efficacy of the approach (i.e., we answer to the question whether our (small) test suite is sufficient to cover all mutants), and check whether it delivers acceptable performance in terms of executions time.

Figure~\ref{fig:bar_plot} reports an histogram for each considered problem. More specifically, each  histogram reports one bar for each mutant. The green (resp. red) color indicates succeeded (resp. failed) tests case executions for each generated mutant. A test succeeds whenever an assertion is satisfied, and fails otherwise.

As expected, the bar corresponding to the original (i.e., bug-free) encoding are only colored in green, which means that no assertion fails on a correct program.
Moreover, some assertion fails in all the mutants (note that the red color is visible in bars corresponding to mutants). 
This means that all the bugs introduced in the encodings for the three problems could be detected.

Given the high \textit{worst case} complexity of checking assertions, and the intrinsic complexity of benchmark problems (recall that HP is NP-complete, FF is NP hard and 2QBF is $\Sigma_2^P$-complete), one might wonder whether this result can be obtained in a reasonable time.

The average execution time employed by \reviewfix{ASP-WIDE} is reported in Table~\ref{tab:time_dist_by_bench}. 
Test cases in HP, FF, and 2QBF were completed in average 0.003s, 0.05s, and 0.026s, respectively. 
This average performance is clearly acceptable, and reveals that the majority of assertion tests is basically instantaneous. 
(A more detailed analysis on these results is provided in the next paragraph).

We remark that this result can be seen as a confirmation of the small-scope hypothesis~\cite{DBLP:conf/kr/OetschPPST12}.
As observed by ~\nCite{DBLP:conf/kr/OetschPPST12},
to detect a bug in an ASP program, it is sufficient to ``analyze programs after grounding them over a small domain''. 
As a consequence, regardless of the high worst case complexity associated with assertion testing tasks, unit testing remains effective in identifying bugs due to the reasonable solving time of state-of-the-art ASP systems.
On the other hand, it is important to take into account the high worst case complexity of assertion testing when designing test cases. Indeed, complexity results tell us that utilizing large instances in a test case can render it ineffective and pointless, as waiting for the results may become unreasonably time-consuming.
A rule of thumb for devising good unit tests is to concentrate in devising smart test cases that run over small problem instances.


\begin{table*}[b!]\footnotesize
    \centering
    \caption{Comparison between \reviewfix{ASP-WIDE} and \reviewfix{ASPIDE-LIKE}.\label{tab:time_dist_by_bench}
    }
    \begin{tabular}{lcrrrrrrrr}
    \hline
    \multicolumn{1}{c}{\multirow{2}{*}{Benchmark}} & \multicolumn{1}{c}{\multirow{2}{*}{\reviewfix{Test Cases}}} & \multicolumn{1}{c}{\multirow{2}{*}{\#}}  & \multicolumn{3}{c}{\reviewfix{ASP-WIDE}} && \multicolumn{3}{c}{\reviewfix{ASPIDE-LIKE}} \\
    \cline{4-6} \cline{8-10}
    \rule{0pt}{12pt}    
    & & & \multicolumn{1}{c}{$\#solved$}& \multicolumn{1}{c}{$avg(T)$} & \multicolumn{1}{c}{$\sigma(T)$} && \multicolumn{1}{c}{$\#solved$} & \multicolumn{1}{c}{$avg(T)$} & \multicolumn{1}{c}{$\sigma(T)$}\\
    \hline
        Hamiltonian Path & \reviewfix{6} & 880 & 880 & 0.003 & 0.018 &  & 565 & \reviewfix{}{216.750} & 286.749\\
        Fast Food & \reviewfix{5} & 230 & 230 & 0.050 & 0.398 &  & 212 & 50.078 & 160.661\\
        2QBF & \reviewfix{6} & 4590 & 4590 & 0.030 & 0.102 &  & 3596 & 137.275 & 244.897\\
    \hline
    \end{tabular}    
\end{table*}


\paragraph{Comparison with ASPIDE.}
Now \reviewfix{ASP-WIDE} and \reviewfix{ASPIDE-LIKE} performance are compared. This is done to measure the improvements that can be obtained by delivering an implementation, like \reviewfix{ASP-WIDE}, that carefully considers the complexity of the task at hand. 
Also an in-depth analysis of the performance of the systems running each different annotation type is performed.

\renewcommand{\imagesSize}{0.8}
\begin{figure*}
    \centering
    \begin{subfigure}[t]{0.49\textwidth}
        \caption{\reviewfix{Assertion: \texttt{@constraintForAll}}\label{fig:assert_cf}}
        \includegraphics[width=0.75\textwidth]{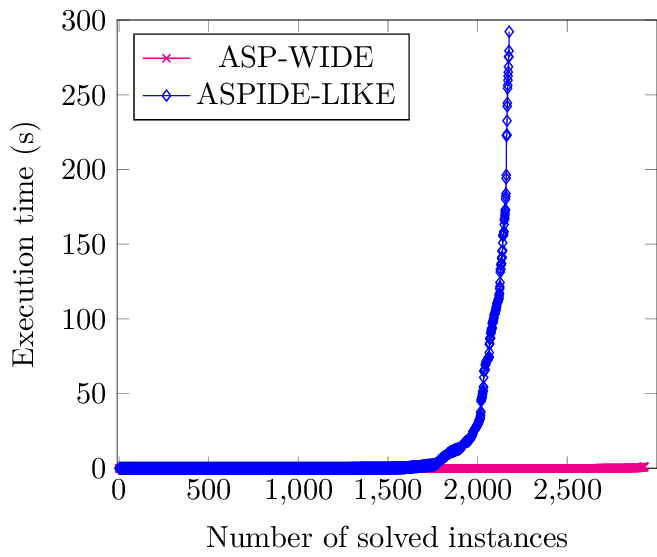}
    \end{subfigure}
    \begin{subfigure}[t]{0.49\textwidth}
        \caption{\reviewfix{Assertion: \texttt{@constraintInAtLeast}}\label{fig:assert_cia}}
        \includegraphics[width=0.75\textwidth]{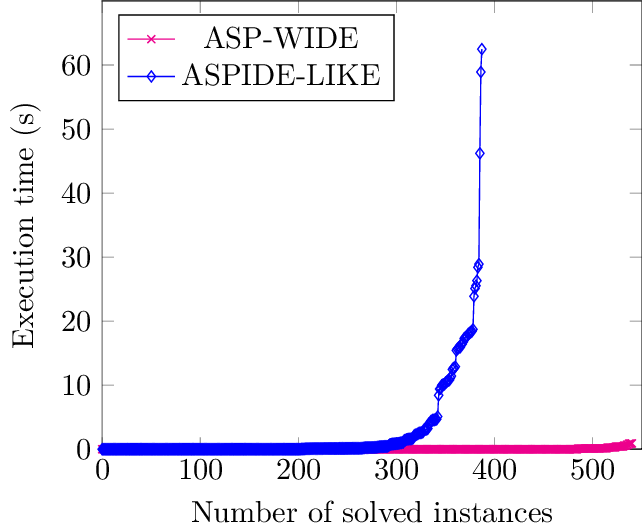}
    \end{subfigure}
    \begin{subfigure}[t]{0.49\textwidth}
        \caption{\reviewfix{Assertion: \texttt{@trueInAll}}\label{fig:assert_tia}}
        \includegraphics[width=0.75\textwidth]{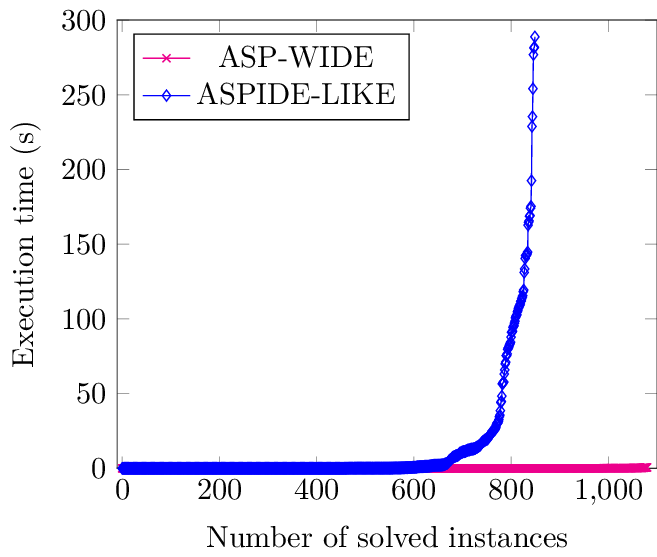}
    \end{subfigure}
    \begin{subfigure}[t]{0.49\textwidth}
        \caption{\reviewfix{Assertion: \texttt{@trueInAtLeast}}\label{fig:assert_tial}}
        \includegraphics[width=0.75\textwidth]{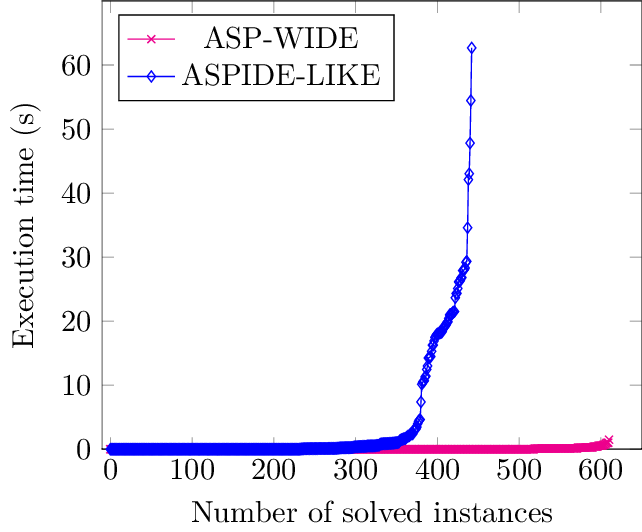}
    \end{subfigure} 
    \begin{subfigure}[t]{0.49\textwidth}
        \caption{\reviewfix{Assertion: \texttt{@trueInExactly}}\label{fig:assert_tie}}
        \includegraphics[width=0.75\textwidth]{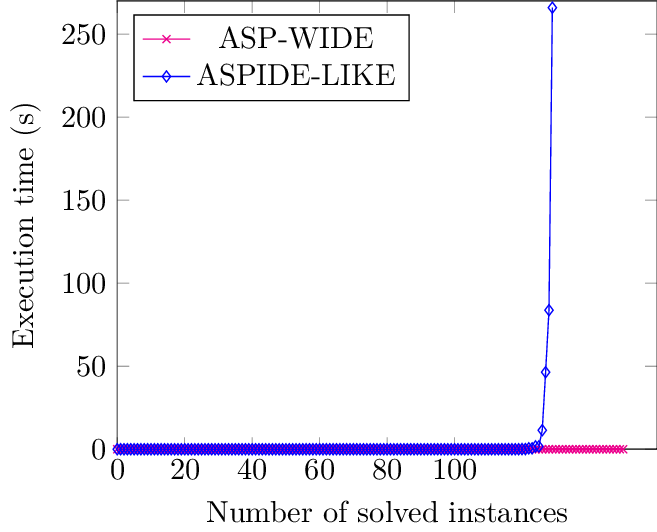}
    \end{subfigure}
    \begin{subfigure}[t]{0.49\textwidth}
        \caption{\reviewfix{Assertion: \texttt{@noAnswerSet}}\label{fig:assert_noAns}}
        \includegraphics[width=0.75\textwidth]{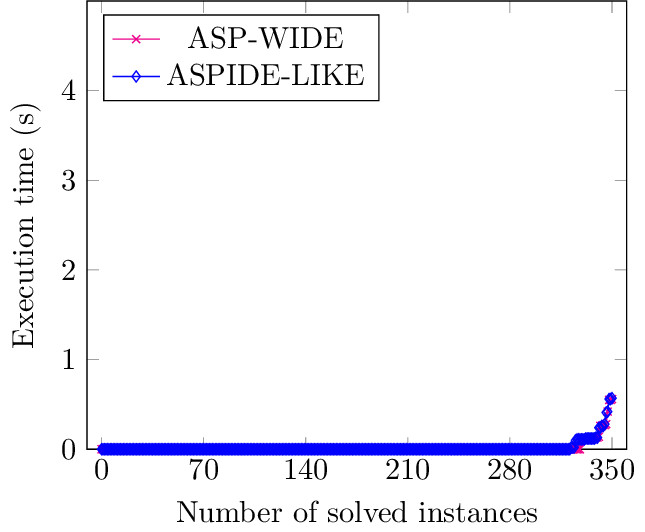}
    \end{subfigure}
    \begin{subfigure}[t]{0.49\textwidth}
        \caption{\reviewfix{Assertion: \texttt{@bestModelCost}}\label{fig:assert_bmc}}
        \includegraphics[width=0.75\textwidth]{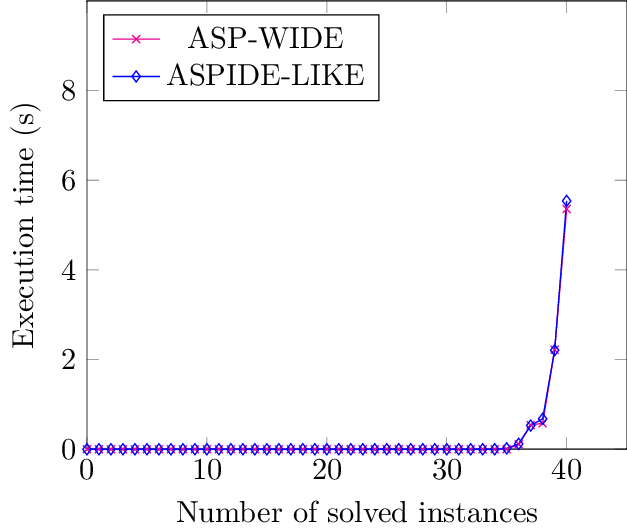}
    \end{subfigure}
    \begin{subfigure}[t]{0.49\textwidth}
        \caption{\reviewfix{Overall performances}\label{fig:overall}}
        \includegraphics[width=0.75\textwidth]{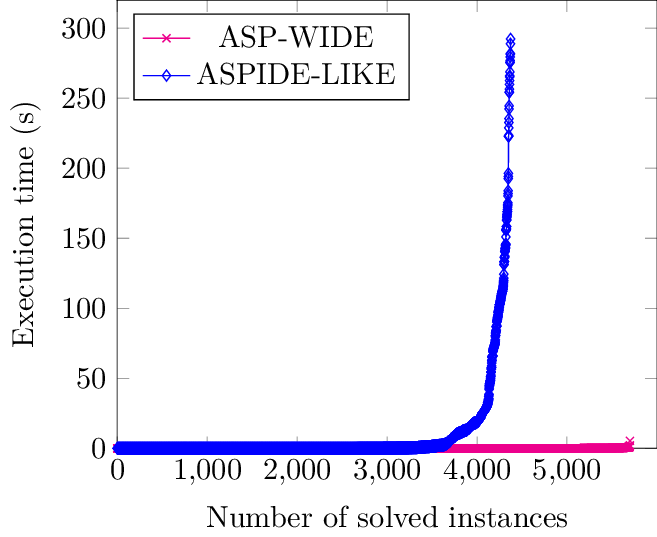}
    \end{subfigure}
    \caption{Assertion-based comparison between \reviewfix{ASP-WIDE} and \reviewfix{ASPIDE-LIKE} methods. \label{fig:cactus_by_assertion}}
\end{figure*}
 
Obtained results are summarized by Figure~\ref{fig:cactus_by_assertion}, and Table~\ref{tab:time_dist_by_bench}. 
In particular, Table~\ref{tab:time_dist_by_bench} reports, for each benchmark and testing method, number of solved instances (\textit{\#solved}), average running time ($avg(T)$), and standard deviation $\sigma(T)$ computed over all the test case executions (column \textit{\#}).
While \reviewfix{ASP-WIDE} consistently generates results within the time limit for all test cases, ASPIDE fails to do so. 
Therefore, average and standard deviation provided in Table~\ref{tab:time_dist_by_bench} take into account the execution time for solved instances and double the time limit (i.e., 600 seconds) for unsolved instances. This performance measure, commonly known as the PAR-2 score, \reviewfix{is widely used in systems comparison~\cite{DBLP:journals/ai/FroleyksHIJS21}.}
From Table~\ref{tab:time_dist_by_bench} it is evident that the \reviewfix{ASP-WIDE} outperforms by orders of magnitude \reviewfix{ASPIDE-LIKE}. 
Nonetheless, the values of the standard deviation suggest that some specific instances are more demanding.

Figure~\ref{fig:cactus_by_assertion} reports the aggregate performance of both \reviewfix{ASP-WIDE} and \reviewfix{ASPIDE-LIKE} in form of cactus plots, one for each assertion type (executed on all encodings). Recall that a line in a cactus plot contains a point $(X,Y)$ if a given method is able to solve $X$ instances within $Y$ seconds.
From Figures~\ref{fig:assert_cf}-\ref{fig:assert_bmc} it emerges that \reviewfix{ASP-WIDE} is very efficient in checking all the assertion types.
\reviewfix{More precisely, ASP-WIDE outperforms \reviewfix{ASPIDE-LIKE}, which often reveals an exponential-like behaviour. Whereas, in the case of \texttt{@noAnswerSet} and \texttt{@bestModelCost}, the two approaches performs equally since the underlying check is the same (cfr. Figure\ref{fig:assert_noAns} and Figure~\ref{fig:assert_bmc})}.
Here, the first answer of the system is sufficient to check the assertions, as the corresponding task has the same computational complexity of checking the existence of an (optimal) answer set.

The overall gap in performance is evident in Figure~\ref{fig:overall}, which compares the two approaches over all executions. 
More in detail, we report that \reviewfix{ASP-WIDE} run 5700 tests within 135 seconds by using, on average, 2.75MB of memory, while, \reviewfix{ASPIDE-LIKE} run 4373 tests in 10 hours by using, on average 94MB of memory. 

\color{black}

\section{Related Work}\label{sec:related}
The first paper approaching the problem of systematic testing of ASP programs is~\cite{DBLP:conf/ecai/JanhunenNOPT10}, where a general framework for structure-based testing of answer set programs, encompassing the definition of test coverage notions for ASP programs, has been proposed. In ~\cite{DBLP:conf/ecai/JanhunenNOPT10} the complexity issues related to coverage problems and the inherent complexity of relevant decision problems were also studied.
An experimental comparison of basic strategies for random testing and structure-based testing of ASP programs has been presented~\cite{DBLP:conf/lpnmr/JanhunenNOPT11}.
Results obtained in this latter indicate that random testing is quite effective in catching errors provided that sufficiently many admissible test inputs are considered.
It has been empirically demonstrated~\cite{DBLP:conf/kr/OetschPPST12} that the small-scope hypothesis of traditional testing holds also in the case of ASP programs. That is, many errors can be found by testing a program w.r.t. test inputs considering a small number of objects (i.e., from a small scope). 
More recently, a new tool for random based testing of ASP programs, called Harvey, has been described~\cite{DBLP:conf/lpnmr/GresslerOT17}.
In Harvey random testing for ASP has been implemented using ASP itself (i.e., both test-input generation and determining test verdicts is done employing ASP). Harvey achieves uniformity of the test-input selection by using XOR streamlining~\cite{DBLP:conf/sat/GomesHSS07} a technique used also in the area of SAT solving~\cite{DBLP:conf/nips/GomesSS06}.
The methods mentioned up to now focus on the problem of generating automatically test suites for ASP programs that are sufficient to identify defects, \textit{after correct programs have been written}.
These tools are thus particularly useful in cases in which one wants to improve an encoding, and use a natural (but less efficient) encoding, to check whether a more complicated (but efficient) one is being developed.
On the other hand, as outlined in the introduction, the goal of unit testing in software development is to drive the implementation towards working software also when \textit{no previous solution exists}. 
Indeed, in TDD, test cases are derived from the requirements even before writing the source code~\cite{DBLP:conf/xpu/FraserBCMNP03,Beck2002}. Nonetheless, automatic test generation can be combined with unit testing, e.g., in case one is evolving a solution to meet some non-functional requirement such as efficient computation.

Focusing on unit testing,  the first implementation of an ASP-specific solution was presented and included in the comprehensive development tool ASPIDE~\cite{DBLP:conf/inap/FebbraroLRR11}. This implementation utilizes rule naming inside of ASP comments in combination with a test definition language for specifying test cases. While rule naming can be accomplished in an  annotation-like manner, which does not interfere with program executability, the specification of test cases required a separate test file and a dedicated syntax~\cite{DBLP:conf/inap/FebbraroLRR11}. 
Since adding meta-information to programs in form of annotations is known from conventional programming languages as C\# and Java, a purely annotation-based test case specification for answer set programs is desirable. 
With \cite{DBLP:journals/tplp/VosKOPT12} a language for annotating answer set programs (called LANA) is presented. Although LANA is not solely devoted to testing, it does address test case definition inside of ASP comments. Despite fully relying on annotations, its implementation (called ASPUnit) requires each unit test to be defined in a separate file, and was never integrated in a development environment (to the best of our knowledge). 
Consequently the desire for a lightweight test definition mechanism that is purely annotation-based and does not necessarily require additional files or external tools that are not included in an environment dedicated to assisted programming remained unfulfilled. 
Moreover, it is important to note that, from a pure syntactic perspective, our assertion language closely resembles the one of ASPIDE, which is also based on Java-like conventions, and uses similar names for the assertions. On the other hand, our language syntax-wise is significantly different from LANA, which follows its own syntactic conventions and can provide the same assertions.

The annotation language presented in this paper, albeit inspired by existing proposals, presents a new syntax that differs from both ASPIDE and LANA proposals, and recalls the well-known annotation style of Java.
Since one of our design goals was to keep it simple while considering all the most important features, our language does not support (by design) some of the ASPIDE-specific options (such as automatic extraction of program modules, and run configuration management), and some of the LANA-specific options (such as pre/post conditions, and signatures).
We discarded those that are: not strictly-related to program testing (e.g., signatures); can be simulated (e.g. pre/post conditions); are implementation specific (run configuration management);  have a not so obvious semantics (automatic expansion of program modules). 
Indeed, automatic expansion of program modules in ASPIDE allows the automatic extension of a block under test with the rules from the original program up to the point that a modularity condition is satisfied, such as the splitting condition~\cite{DBLP:conf/iclp/LifschitzT94} or the more precise conditions of~\cite{DBLP:journals/jair/JanhunenOTW09}. In our experience this feature augments the program under test in a way that is not obvious to the programmer, thus we decided to discard this feature. 
Alternative ways of verifying the correctness of programs with input and output have been recently proposed~\cite{DBLP:journals/tplp/FandinnoLLS20}, that could be considered for integration in our framework.

We remark that our approach allows (as first suggested by \reviewfix{\citeANP{DBLP:journals/tplp/VosKOPT12}~\citeyear{DBLP:journals/tplp/VosKOPT12}}) to inline tests with code. This choice brings advantages, e.g., makes it easy to pack and distribute ASP programs, that are often contained in a single file, with  tests. 
However, the programmer can decide to keep tests and code in separate files, which might be convenient in some cases (e.g., the ASP code is divided in several files); thus our approach can deliver the maximum flexibility in the organization of a project.


Finally, we observe that our unit testing language has been conceived for ASP-Core-2, nonetheless, it can be applied -almost as it is-  to any extension of ASP, such as the richer languages supported by Clingo~\cite{DBLP:journals/tplp/GebserKKS19} and DLV 2.0~\cite{DBLP:conf/lpnmr/AlvianoCDFLPRVZ17}, DLVHEX~\cite{DBLP:journals/ki/EiterGIKRSW18}, ASPMT~\cite{DBLP:journals/ai/BartholomewL19,DBLP:conf/kr/ShenL18}, CASP~\cite{DBLP:journals/amai/MellarkodGZ08,DBLP:journals/tplp/BalducciniL17,DBLP:journals/tplp/BanbaraKOS17}, and SPARC~\cite{DBLP:conf/lpnmr/BalaiGZ13}. Indeed, annotations (in comments) do not interfere with the specifications. Moreover, our approach could be complemented with techniques for rushing and strolling among answer sets~\cite{DBLP:conf/aaai/FichteGR22}. 


\section{Conclusion}\label{sec:conclusion}
Unit testing frameworks are nowadays considered best practice in all modern software development processes. 
The development of ASP applications can be accelerated by resorting to unit testing frameworks, as it happens for all known programming languages.
In this paper we revisit unit testing in ASP by proposing a new unit test language that unifies the strengths of previous approaches.
The new language allows the development of test cases inline with ASP code, keeps the style of expressing test case conditions from ASPIDE, and is annotation-based as LANA. 
Moreover it features a refreshed syntax that is nearer to the JUnit framework, and should look more familiar to developers that are accustomed to XUnit style languages.
Importantly, the new unit testing language is implemented in a novel web-based development  environment for ASP, which supports test driven development of ASP programs.
Another contribution of the paper is a complete identification of the computational complexity of the tasks associated to unit testing of ASP programs. To the best of our knowledge this overview was never done in the literature, and contains both known and novel results.

Despite the high worst case complexity of testing ASP programs, thanks to the small-scope hypothesis, one can define effective test suites with good performance in practice. 
Moreover, the experiment reported in the paper reveals the importance of devising a complexity-aware implementation, as \reviewfix{ASP-WIDE} outperforms the more naive approach of ASPIDE~\cite{DBLP:journals/corr/Febbraro11}. 
\color{black}

As far as future work is concerned, we are improving the \reviewfix{ASP-WIDE} environment by implementing additional features, such as quick fixes, code templates, support for debuggers that are available in more mature IDEs for ASP~\cite{DBLP:conf/inap/FebbraroLRR11,DBLP:journals/tplp/BusoniuOPST13}; moreover we are evaluating the possibility of resorting to a dedicated answer set counting system~\cite{DBLP:journals/corr/FichteHMW16} to implement test case conditions that are $\mathsf{PP}$-complete and $\mathsf{C}\cdot\mathsf{coNP}$-complete.


\bibliographystyle{acmtrans}
\bibliography{bibtex}

\label{lastpage}
\end{document}